# Hydraulic Fracture Propagation in Naturally Fractured Reservoirs: Complex Fracture or Fracture Networks*

HanYi Wang, Petroleum & Geosystems Engineering Department, The University of Texas at Austin

**Abstract**

All reservoirs are fractured to some degree. Depending on the density, dimension, orientation and the cementation of natural fractures and the location where the hydraulic fracturing is done, pre-existing natural fractures can impact hydraulic fracture propagation and the associated flow capacity. Understanding the interactions between hydraulic fracture and natural fractures is crucial in estimating fracture complexity, stimulated reservoir volume (SRV), drained reservoir volume (DRV) and completion efficiency. However, what hydraulic fracture looks like in the subsurface, especially in unconventional reservoirs, remain elusive, and many times, field observations contradict our common beliefs. In this study, a global cohesive zone model is presented to investigate hydraulic propagation in naturally fractured reservoirs, along with a comprehensive discussion on hydraulic fracture propagation behaviors in naturally fractured reservoirs. The results indicate that in naturally fractured reservoirs, hydraulic fracture can turn, kink, branch and coalesce, and the fracture propagation path is quite complex, but it does not necessarily mean that fracture networks can be created, even under low horizontal stress difference, because of strong stress shadow effect and flow-resistance dependent fluid distribution. Perhaps, 'complex fracture', rather than 'fracture networks', is the norm in most unconventional reservoirs.

## 1. Introduction

Since its introduction, hydraulic fracturing has been established as the premier production enhancement procedure in the petroleum industry and has continued to overwhelmingly dominate low-permeability reservoirs as one of the most important field development operations. In very high permeability reservoirs, hydraulic fractures have a dual purpose: to stimulate the well and to provide sand control. In moderate permeability reservoirs, the fracture accelerates production without impacting the well reserves. In low permeability reservoirs, hydraulic fracture contributes both to well productivity and to the well reserves, because in such reservoirs the well would not produce an economic rate without the hydraulic fracture (Economides et al. 2013). In homogenous reservoir with planar fracture geometry, there exists an optimum fracture dimension that would provide maximum reservoir performance for a given amount of proppant (Wei and Economides et al. 2005). So predicting hydraulic fracture dimension by simulating hydraulic fracture propagation is crucial in hydraulic fracturing and completion design.

Geological and field studies have shown that a significant portion of hydrocarbon resources are situated in low permeability naturally fractured formations. For instance, core and outcrop data study from 18 shale plays in North America reveals that natural fractures are commonplace and are concentrated where clay content is low (Gale et al. 2014). Field study (Raterman et al. 2017) by coring rock sample at Eagle Ford reveals that natural fractures are abundant and hydraulic fractures are complex. Studies from a major shale play from Sichuan Basin in China show the existence of both interlaminated and structural fractures, and some fractures are partially or completely filled with calcite (Liang et al. 2014; Zeng et al. 2016). Besides shales, natural fractures are also abundant in many sandstone reservoirs, such as the middle member of Bakken Formation (Pitman et al. 2001), the Mesaverde Group in the Piceance Basin (Fall et al. 2015), the Lower Jurassic Ahe Formation of Tarim Basin (Ju et al. 2018). Natural fractures are more common in carbonate rocks, and in fact, most carbonate reservoirs are naturally fractured (Garland et al. 2012). Poorly sealed natural fractures can interact heavily with the hydraulic fractures during the injection treatments, serving as preferential paths for the growth of complex fracture network, as is evident from microseismic monitoring (Cipolla et al. 2011; Zakhour et al. 2015), or arrest fracture growth along certain directions (Olson et al. 2012; Gu et al. 2012). Rate transient analysis (RTA) from production data also indicates that the productive fractures are indeed complex in many reservoirs (Acuña 2016). Understanding how hydraulic fracture interacts with natural fractures under different scenarios is crucial for well placement and hydraulic fracture design in naturally fracture reservoirs, because all the factors (such as fracture dimension, fracture connectivity, fracture orientation, proppant transport and stress-dependent fracture conductivity) that determine the production rate and ultimate recovery hinge on it. Besides the extraction of hydrocarbons, hydraulic fracturing in naturally fractured reservoirs also has significant implications for geothermal reservoir development to enhance heat recovery (Legarth et al. 2005; McClure and Horne 2014).

The existence of natural fractures adds tremendous challenges to the hydraulic fracturing design. Since the introduction of the concept of stimulated reservoir volume (SRV) by Mayerhofer et al. (2010), "complex network structures" has become a common postulation when we envision what hydraulic fracture really like in naturally fractured unconventional reservoirs. Besides indirect measurements and observations, such as microseismic monitoring, tiltmeter measurement and well interference analysis, that can be used to infer the upper bound of fracture dimension, it is extremely difficult, if not impossible, to quantify hydraulic fracture typology on a field scale. Numerical modeling and simulation provide us an alternative to gauge the behavior of hydraulic fracture propagation in naturally fractured reservoir that cannot be substituted by small-scale laboratory experiment, where it is difficult to replicate in-situ conditions with extremely high confining stress, in addition, the minimum stress tends to elevate (i.e., stress anisotropy can be much lower than designed or even disappear) during fracture propagation because of fixed displacement at the rock sample boundary (Malhotra et al. 2018), and the boundary effect can't be fully eliminated even one can maintain a constant far-field stress field during fracture propagation, and on top of that, the impact of flow-resistance dependent fluid distribution is not able to fully manifest itself in small-scale experiment, thus, laboratory experiment has the tendency to create hydraulic fracture patterns that are more complex than what is happening in the subsurface.

Currently, a number of extraordinary works (Dahi-Taleghani and Olson 2011; McClure and Horne 2013; Sesetty and Ghassemi 2017; Shrivastava and Sharma 2018; Weng et al. 2011) have investigated how hydraulic fracture interacts with natural fractures based on the displacement discontinuity method (DDM). DDM is a type of boundary integral method where the fundamental solution is derived from analytical solutions to the problem of a constant discontinuity in displacement on a finite segment in an infinite or semi-infinite medium. The advantages of the DDM are that it only requires discretization at the boundaries of the fracture surface; in addition, unlike the finite element method (FEM) or finite volume method (FVM), it does not need re-discretization for any new crack, and hence, is generally more computationally efficient because much less degree of freedom is needed. Because the underlying assumptions of DDM are based on linear elastic fracture mechanics (LEFM), so it is only applicable to brittle rocks, which limits its applications in quasi-brittle/ ductile rocks. Laboratory experiments (Sone and Zoback 2013; Hull et al. 2015) on shale samples from a variety of shale plays in U.S found that the ductility of shale sample is strongly correlated to its organic and clay content. Furthermore, sedimentary rocks tend to transit from brittle to ductile under high confining pressure (Wong and Baud 2012), so it may not be appropriate to use DDM in deep reservoirs or shale plays with a relatively high content of clay or kerogen. In addition, DDM also assumes homogenous properties, so it has difficulties tackling some challenging cases, such as various rock properties along the horizontal wellbore, multi-layer formations with different mechanic properties and strength or near-wellbore fracture modeling where casing and cement need to be included. On top of that, to simulate the interaction between hydraulic fracture and natural fractures, most DDM models require a pre-defined crossing criterion that derived from the assumptions of plain strain condition, infinite elastic domain and local Mohr-Coulomb law (Renshaw and Pollard 1995; Gu and Weng 2010), thus it is challenging to apply DDM in naturally fractured reservoirs in a fully coupled manner.

In this study, the cohesive zone method (CZM) is implemented to model hydraulic fracture propagation, which does not have the above limitations and is a promising approach to simulate complex fracture propagation and interaction. Barenblatt (1959; 1962) first introduced the conception of cohesive zone to model crack propagation in brittle materials. Dugdale (1960) proposed a fracture process zone to investigate fracture propagation in ductile materials with small-scale of plasticity. Mokryakov (2011) presented a cohesive zone model for fracturing soft rock, which leads to more accurate fitting of pressure log. Wang et al. (2016) developed a hydraulic fracture model for both poroelastic and poroplastic formations with the cohesive zone method, where the effects of plastic deformation near the fracture tip and inside the reservoir are considered. Guo et al (2017) proposed a cohesive zone model to investigate hydraulic fracture in a layered reservoir to study the influence of geologic and fracture execution parameters. Wang (2015; 2016) combined extended finite element method (XFEM) and cohesive zone method to study fracture re-orientation from unfavorable perforations, and fracture interference and coalescence from multi-stage fracturing. Manchanda et al (2017) proposed a 3D cohesive zone model for multiple hydraulic fracture propagation where the heterogeneity of rock properties is included. Baykin and Golovin (2018) presented a fully coupled poroelastic cohesive model to investigate the effect of permeability contrast on hydraulic fracture propagation. Despite the successful implementations of cohesive zone method to model planar and non-planar hydraulic fracture propagation, the work on modeling of hydraulic fracture interacts with natural fractures using cohesive zone method is very limited. Guo et al. (2015) and Chen et al. (2017) investigated a hydraulic fracture interacting with a single pre-existing natural fracture using cohesive zone approach, Dahi Taleghani et al. (2018) extended their models and introduced fracture networks into their simulation, but all these studies require predefining a fracture propagation path within the intact rock domain, and fractures can only propagate along the predefined path and weak planes, which hinder the generality of their models when simulating complex fracture growth, where the fracture propagation path itself is solution dependent.

The structure of this article is as follows. First, a global cohesive zone model is presented, which is capable of simulating hydraulic fracture propagation and interaction with natural fractures. Next, synthetic simulation cases are investigated, including ones with randomly generated conjugate natural fracture sets. Then, a comprehensive discussion on what hydraulic fracture really looks like in naturally fractured reservoirs is introduced and the implications of filed observations are also examined. Finally, concluding remarks are presented.

## 2. Global Cohesive Zone Model

For a Newtonian fluid with viscosity μ, the flow inside fracture can be approximated by flow between parallel plates, and the local flow rate $\boldsymbol{q_f}$ can be determined by the pressure gradient and the local fracture width (Boone and Ingraffea, 1990):

$$\boldsymbol{q_f} = -\frac{w^3}{12\mu}\nabla p_f, \quad \quad \quad \quad \quad \quad \quad \quad \quad \quad \quad \quad (1)$$

where $w$ is the local fracture width, $p_f$ is the fluid pressure inside the fracture. The conservation of the fluid mass inside the fracture can be described by the lubrication equation:

$$\nabla \boldsymbol{q_f} - \frac{\partial w}{\partial t} + q_l = 0, \quad \quad (2)$$

where $q_l$ is the local fluid loss in rock formation from leak-off per unit fracture surface area. Pressure dependent leak-off model is used in this study to describe the normal flow from fracture into surrounding formations:

$$q_l = c_l(p_f - p_m), \quad \quad (3)$$

where $p_m$ is pore pressure in the formation adjacent to the fracture and $c_l$ is pressure-dependent leak-off coefficient, which is a constant value. Darcy's Law is used to describe fluid diffusion in the porous media of rock matrix:

$$\boldsymbol{q_m} = -\frac{\boldsymbol{k}}{\mu}\nabla p_m, \quad \quad (4)$$

where $\boldsymbol{k}$ is formation permeability tensor and $\boldsymbol{q_m}$ is the fluid flux velocity vector in the porous media. In fluid-filled porous media, the total stresses $\sigma_{i,j}$ are related to the effective stresses $\sigma'_{i,j}$ through a poroelastic constant $\alpha$ and pore pressure $p_w$ (Biot 1941):

$$\sigma_{i,j} = \sigma'_{i,j} + \alpha p_w \quad \quad (5)$$

In this study, the value of poroelastic constant $\alpha$ is assumed to be 0.7. The equilibrium equation of poroelasticity in the form of virtual work principle for the volume under its current deformation at time $t$ can be written as:

$$\int_V (\sigma' + p_w \mathbf{I}) : \delta\varepsilon dV = \int_S \boldsymbol{t}\delta v dS + \int_V \boldsymbol{f}\delta v dV \quad \quad (6)$$

where $\mathbf{I}$ is the unit matrix, $\boldsymbol{t}$ is the surface traction per unit area, $\boldsymbol{f}$ is the body force per unit volume. $\sigma'$ is the effective stress and $\delta$ denotes virtual component, and $\delta\varepsilon$ is the virtual rate of deformation, $\delta v$ is virtual displacement. Equating the change rate of the total mass of fluid in the control volume V to the mass velocity of fluid crossing the surface S gives the fluid mass continuity equation in the following form:

$$\frac{d}{dt}\left(\int_V \rho_w n_w dV\right) + \int_S \rho_w n_w \boldsymbol{n}\, \boldsymbol{v_w} dS = 0 \quad \quad (7)$$

where $\boldsymbol{n}$ is the outward normal to the surface S, $\boldsymbol{v_w}$ is the fluid velocity, $\rho_w$ is the mass density of the liquid and $n_w$ is the porosity of the porous medium. The criterion for fracture initiation and propagation is based on cohesive zone method. The traction force T and the displacement $\delta$ across a pair of cohesive surfaces is related by a cohesive potential function :

$$\mathrm{T} = \frac{\partial G}{\partial \delta} \quad \quad (8)$$

In this study, a bilinear cohesive law (Tomar *et al.*, 2004) is used to describe the relationship between traction force and displacement. This law assumes that the cohesive surfaces exhibit reversible linear elastic behavior until the traction reaches the cohesive strength at a damage initiation displacement of $\delta_0$. Beyond $\delta_0$, the traction reduces linearly to zero until complete failure at a displacement of $\delta_f$. The area under the traction-displacement curve equals the critical fracture energy $G^c$, which is the work needed to create a unit area of fully developed fracture. For elastic solids this energy is related to the rock fracture toughness $\mathrm{K}_{IC}$ through Young's modulus E and Poisson's ratio $\nu$ (Kanninen and Popelar, 1985):

$$G^c = \frac{\mathrm{K}_{IC}^2}{E}(1 - \nu^2) \quad \quad (9)$$

Damage is assumed to initiate when normal or shear strength is reached. This criterion can be represented as

$$\left\{\frac{\langle\sigma_n\rangle}{\sigma_n^0}\right\}^2 + \left\{\frac{\sigma_s}{\sigma_s^0}\right\}^2 + \left\{\frac{\sigma_t}{\sigma_t^0}\right\}^2 = 1 \qquad (10)$$

where $\sigma_n^0, \sigma_s^0, \sigma_t^0$ represent the normal and two shear strength of intact cohesive surfaces. and $\sigma_n, \sigma_s, \sigma_t$ refer to the normal, the first, and the second shear stress components; The symbol $\langle\ \rangle$ used in the above equation represents the Macaulay bracket to signify that a pure compressive deformation or stress state does not initiate damage. The stress components of the traction-separation model are affected by the damage according to

$$\boldsymbol{\sigma} = \begin{cases} (1-D)\bar{\boldsymbol{\sigma}} & \text{damage initated} \\ \bar{\boldsymbol{\sigma}} & \text{no damage occurs} \end{cases} \qquad (11)$$

where $\boldsymbol{\sigma}$ are stress components, $\bar{\boldsymbol{\sigma}}$ are the stress components predicted by the elastic traction-separation behavior for the current displacement without damage. D is a scalar damage variable that represents the overall damage in the material. D monotonically increases from 0 to 1 upon after the initiation of damage with further loading. The evolution of the damage variable, D, is determined by traction-separation law. In this study, a bilinear traction-separation law is used and displacement of complete failure $\delta_f$ inside a cohesive zone can be determined by:

$$\delta_f = \frac{2G_I^c}{T_{Max}} = \frac{2K_{IC}^2(1-\nu^2)}{ET_{Max}} \qquad (12)$$

where $T_{Max}$ is the tensile or shear strength. The stress-displacement relation (i.e., σ v.s δ curve) before failure corresponds to linear elastic deformation as follows:

$$\sigma = \frac{\delta T_{Max}}{\delta_0} = \frac{\delta E}{d_z} \qquad (13)$$

where $d_z$ is the initial thickness of cohesive surfaces that is calculated as

$$d_z = \frac{9\pi E G_{IC}}{32(1-\nu^2)T_{Max}^2} \qquad (14)$$

the post-peak softening regime after failure is given by the following:

$$\sigma = T_{Max}\left[1 - \frac{(\delta - \delta_0)}{(\delta_f - \delta_0)}\right] \qquad (15)$$

and the displacement of $\delta_f$ at complete failure is determined by:

$$\delta_f = \frac{2G^c}{T_{Max}} \qquad (16)$$

The damage variable, D, is then calculated as

$$D = \frac{\delta_m^f(\delta_m^{max} - \delta_m^0)}{\delta_m^{max}(\delta_m^f - \delta_m^0)} \qquad (17)$$

where $\delta_m$ is effective displacement, defined as

$$\delta_m = \sqrt{\langle\delta_n\rangle^2 + \delta_s^2 + \delta_t^2} \qquad (18)$$

where $\delta_n, \delta_s, \delta_t$ refer to the normal, the first, and the second shear displacement components. Bilinear traction-separation law is simple to implement because it only requires the determination of tensile/shear strength, critical fracture energy, Young's modulus and Poisson's ratio. Other more sophisticated traction-separation law is also available, but requires laboratory data from well-designed experiments, such as three points bending test (Lee et al. 2010), semi-circular bending test (Dahi-Taleghani et al. 2018). The coulomb friction law is applied to the failed but contacting surfaces, where the shear slippage occurs when

$$|\tau_s| = \begin{cases} \eta\sigma_n & (\eta\sigma_n \leq \tau_{max}) \\ \tau_{max} & (\eta\sigma_n > \tau_{max}) \end{cases} \qquad (19)$$

where $\eta$ is the coefficient of friction, $\sigma_n$ is the normal compressive stress, $\tau_s$ is the frictional shear stress and $\tau_{max}$ is the shear stress limit on the contacting surfaces.

Finite element method (Zienkiewicz and Taylor 2015) is used to discretize the above equations in the simulation domain. Zero thickness cohesive zones are embedded along element boundaries and fracture can initiate and propagation along all the boundaries, so that the propagation, branching, merging, and intersection of fractures can be captured. Natural fractures are represented by cohesive zones which have much lower tensile strength, shear strength and critical fracture energy than the intact rock mass, and the direction of the propagation is determined by the local stress state and the planes of weakness around the fracture tip.

## 3. Model Verification and Comparison

In this section, the global cohesive zone model is verified and compared against analytic solutions and empirical criteria from experimental studies. First, analytical solutions from KGD model (Khristianovich and Zheltov 1955; Geertsma and De Klerk 1969) are compared with the simulation results. **Table 1** shows all the input parameters for the KDG verification simulation. The combination of input parameters is designed such that the simulation domain is much bigger than the fracture aperture and length to avoid boundary effect. The value of permeability is chosen to minimize the effect of poroelastic adjacent to the fracture. Cohesive properties are selected to ensure a small cohesive zone ahead of the fracture tip relative to the size of the fracture, and the rock behaves linear-elastically. **Fig1** shows the simulation results. The comparison indicates that the simulation results of the proposed model match very well with the analytical solution, except for the net pressure in the early-stage of injection; this is due to the fact that the analytical solution does not account for the pressure build-up period before breakdown.

**Table 1** Input parameters for KDG problem

| Input Parameters | Value |
|---|---|
| Elastic modulus | 20 GPa |
| Poisson's ratio | 0.25 |
| Fluid viscosity | 1 cp |
| Tensile strength of intact rock | 2 MPa |
| Formation effective permeability | 1 md |
| Injection rate per unit reservoir thickness | 0.0005 m$^3$/s |
| Specific weight of fluid | 9.8 kN/m$^3$ |
| Initial pore pressure | 20 MPa |
| Maximum horizontal stress | 42 MPa |
| Minimum horizontal stress | 37 MPa |
| Vertical stress | 65 MPa |
| Critical fracture energy | 100 J/m$^2$ |
| Pressure dependent leak-off coefficient | 4E-14 m$^3$/s/Pa |
| Porosity | 0.2 |

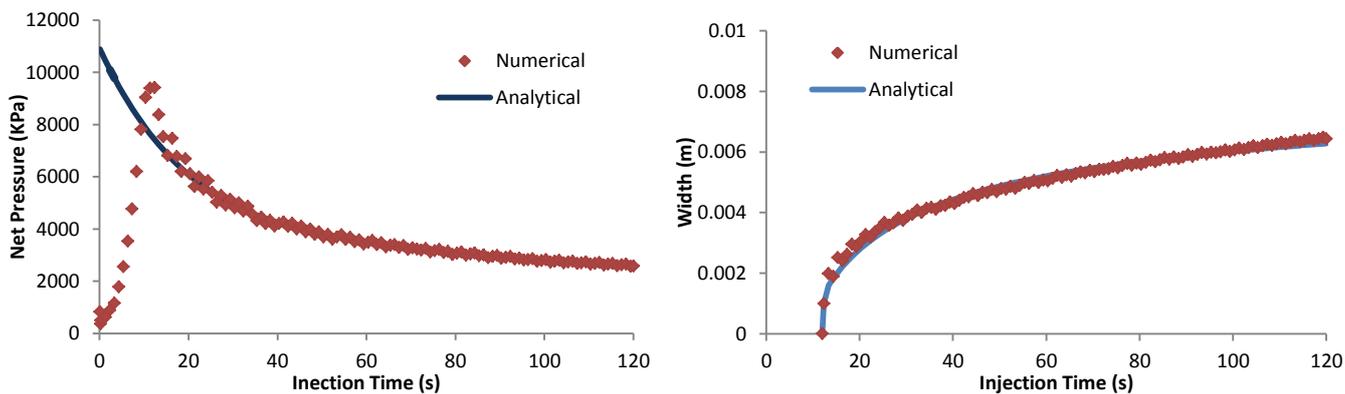

**Fig.1. Net pressure and width at wellbore in elastic formation during injection**

Next, The interactions between a hydraulic fracture and a natural fracture are simulated, as shown in **Fig.2**. There are two possible outcomes from this interaction: one is slippage or arrest, and the other is crossing (Gu et al. 2012). In the first case, it is assumed that the natural fracture intersects the hydraulic fracture path at a 45-degree angle and the natural fracture serves as a weak plane within the intact rock. All the input parameters for the intact rock and the natural fracture are provided in **Table 2** and the horizontal stress difference (HSD) varies between 0 to 8 MPa.

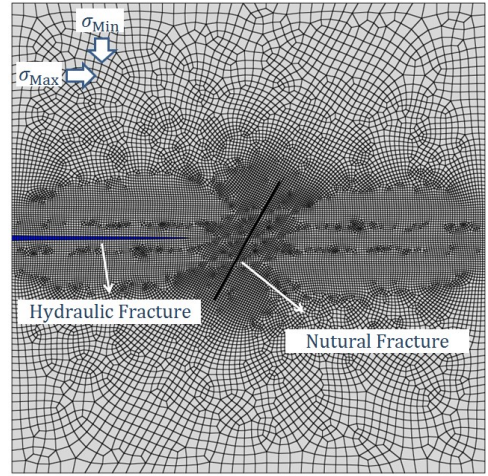

**Fig.2.** Illustration of a hydraulic fracture approaching a pre-existing natural fracture within discretized simulation domain

**Table 2** Input parameters for simulating a hydraulic fracture interacts with a pre-existing natural fracture

| Input Parameters | Value |
| --- | --- |
| Elastic modulus | 15 GPa |
| Poisson's ratio | 0.25 |
| Fluid viscosity | 1 cp |
| Tensile strength of intact rock | 6 MPa |
| Shear strength of intact rock | 20 MPa |
| Tensile strength of natural fracture | 0.2 MPa |
| Shear strength of natural fracture | 1 MPa |
| Formation effective permeability | 1 md |
| Injection rate | 0.001 m$^3$/s |
| Specific weight of injection fluid | 9.8 kN/m$^3$ |
| Initial pore pressure | 5 MPa |
| Tensile critical fracture energy for intact rock | 100 J/m$^2$ |
| Tensile critical fracture energy for natural fracture | 10 J/m$^2$ |
| Shear critical fracture energy for intact rock | 4500 J/m$^2$ |
| Shear critical fracture energy for natural fracture | 450 J/m$^2$ |
| Pressure dependent leak-off coefficient | 1E-14 m$^3$/s/Pa |
| Porosity | 0.1 |
| Friction coefficient | 0.615 |

**Fig.3** shows two examples of the simulation results at different HSD. As can be seen, when HSD is zero, the hydraulic fracture is arrested by the natural fracture at the end of the simulation, and when HSD is 4 MPa, the hydraulic fracture propagates cross the natural fracture and maintains its propagation direction. Blanton (1982) conducted laboratory hydraulic-fracture experiments in pre-fractured hydrostone blocks under tri-axial stresses and proposed a criterion to predict the interaction between hydraulic fracture and natural fracture. According to the criterion, a curve can be constructed by plotting the approaching angle (between the hydraulic and natural fracture) and HSD for constant fracture toughness, as shown in **Fig.4**. The hydraulic fracture will be diverted by the natural fracture at any point below this curve. On the other hand, the hydraulic fracture will cross the natural fracture at any point above this curve. To verify the presented model, additional cases with different approaching angle and HSD are simulated and the results are plotted in Fig.4. As can be seen, the simulation results agree very well with what the criterion predicts.

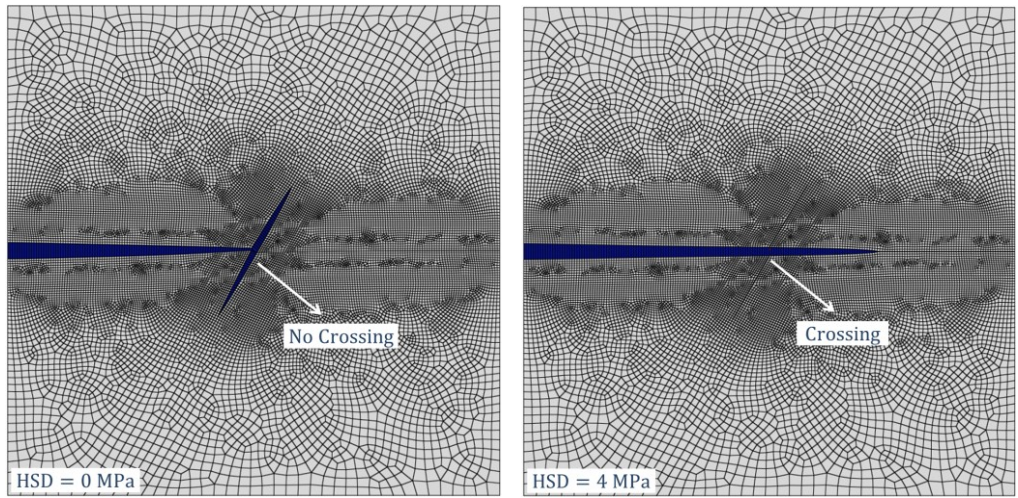

**Fig.3.** Simulation results of a hydraulic fracture interacting with a pre-existing natural fracture at different HSD

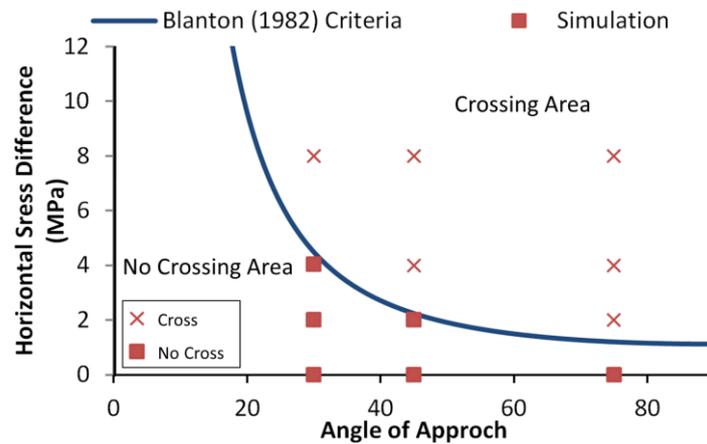

**Fig.4.** Comparison between simulation results and criterion prediction

## 4. Fracture Propagation in Naturally Fractured Reservoirs

Depending on the natural fracture properties and the in-situ stress state, the presence of natural fractures may significantly impact the propagation of hydraulic fractures and the overall efficiency of hydraulic fracture treatments. On one hand, the stimulated natural fractures, if well-connected, can enlarge the reservoir contact area, expand the fluid flow path and enhance productivity. On the other hand, the reactivated natural fracture and weak planes may divert fluid flow from the main hydraulic fracture channel and lead to an unreasonably high proppant concentration and early screenouts. So in order to optimize well placement and hydraulic fracture design in naturally fractured reservoirs, it is crucial to efficiently simulate fracture propagation in these reservoirs. In this section, simulation cases of hydraulic fracture propagating in naturally fractured reservoirs using the proposed model are presented and discussed.

Many naturally fractured reservoirs have more than one natural fracture set (i.e., a fracture set that has one dominant orientation) and reservoirs that contain two natural fracture sets with different orientations are very common. (Dezayes et al. 2010; Hanks et al. 1997; King et al. 2008; Zhang and Li, 2016). Normally, the initial natural fracture set is related to tectonic stresses and develops parallel to the compression direction and the second fracture set results from visco-elastic effects or from orthogonal or sub-orthogonal loadings (Hanks et al. 1997). **Fig. 5** shows the setup of a 2D reservoir model with two sets of pre-existing natural fractures. Because natural fractures are formed within geological time scale and the formation itself may have gone through multiple tectonic events, the distribution and orientation of the natural fractures do not necessarily relate to the current in-situ stresses. For the model presented in this section, the angle between the direction of minimum horizontal stress and that of the two natural fracture sets are 30 and 150 degrees, respectively (i.e., conjugate fractures). The injection well is located at the center of the simulation domain and the perforations are aligned with the direction of maximum horizontal stress, so that the hydraulic fracture is initiated against the minimum horizontal stress. All the input parameters are provided in **Table 3**, which assumes that the two natural fracture sets have the same properties, and the strength of the natural fracture bond is only 1/5 of the strength of the intergranular bond of intact rock.

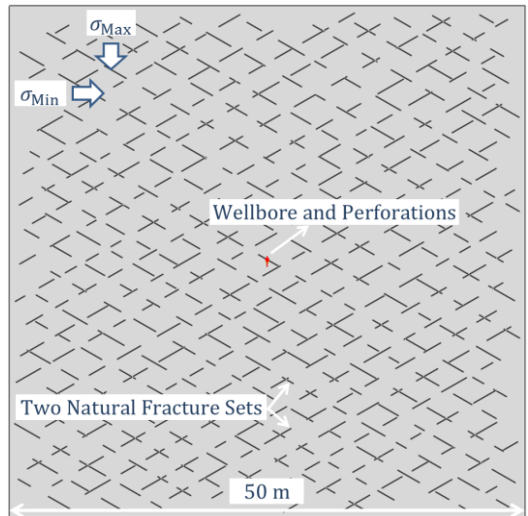

**Fig.5. Simulation setup for reservoirs with two pre-existing natural fracture sets**

Table 3  Input parameters for simulating hydraulic fracture propagates in naturally fractured reservoirs

| Input Parameters | Value |
|---|---|
| Young's modulus | 10 GPa |
| Poisson's ratio | 0.25 |
| Fluid viscosity | 1 cp |
| Tensile strength of intact rock | 5 MPa |
| Shear strength of intact rock | 20 MPa |
| Tensile strength of natural fracture | 1 MPa |
| Shear strength of natural fracture | 4 MPa |
| Formation effective permeability | 1 md |
| Injection rate | 0.01 m$^3$/s |
| Specific weight of injection fluid | 9.8 kN/m$^3$ |
| Initial pore pressure | 10 MPa |
| Tensile critical fracture energy for intact rock | 100 J/m$^2$ |
| Tensile critical fracture energy for natural fracture | 20 J/m$^2$ |
| Shear critical fracture energy for intact rock | 4500 J/m$^2$ |
| Shear critical fracture energy for natural fracture | 900 J/m$^2$ |
| Pressure dependent leak-off coefficient | 1E-14 m$^3$/s/Pa |
| Porosity | 0.1 |
| Friction coefficient | 0.615 |

First, let's examine a case where the horizontal stress difference is large (i.e., HSD=10 MPa). The simulated hydraulic fracture propagation path and the associated fracture width profile are shown in **Fig.6**. As expected, because of the existence of natural fractures, the hydraulic fracture propagation path is not planar. There are two competing forces that govern the general trend of hydraulic fracture propagation path: one is the propensity to propagate along the direction of the weak planes resulting from the stimulated natural fracture sets, and the other is the tendency to propagate along the direction of maximum principal stress and open against the minimum principal stress. More often than not, the directions of natural fracture sets are not aligned with in-situ principal stresses, so the fracture propagation has to balance these two forces and minimize the fracturing energy to find the path that has the least resistance, that's why we see the hydraulic fracture propagates along a zig-zag path. In this particular case, because the HSD is large enough, so the overall hydraulic fracture propagation path is perpendicular to the direction of the minimum horizontal stress. If we look at the fracture width profile, we can also notice that the fracture width is larger close to the wellbore and smaller when the fracture segment is not well-aligned with the maximum horizontal stress direction. Note that the scale of deformation is enlarged in all the following presented figures to better illustrate the fracture propagation path. **Fig.7** shows the total displacement contour within the simulated domain after 50 seconds of injection, which is an indication of how far the rock can feel the "pushes" of open fractures from an initial state. A very important observation is that the formation can have millimeter scale deformation even at 15 meters away from the hydraulic fracture, which can lead to very strong stress shadow effect, and it has significant implications for competing fracture growth and complex fracture typology. The phenomenon of stress shadow effect will be thoroughly discussed in later content.

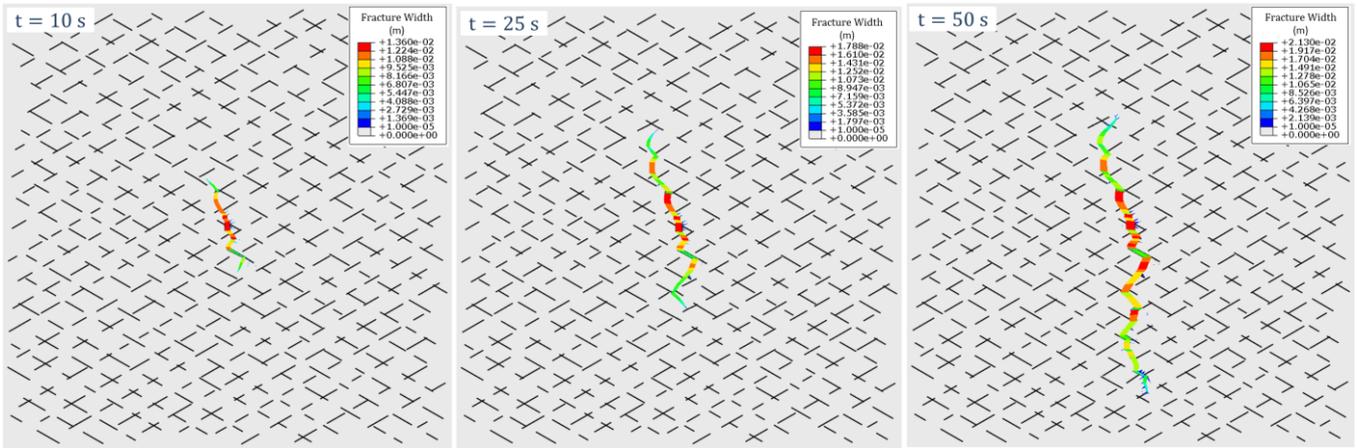

**Fig.6. Simulated hydraulic fracture propagation path and width profile at different injection time, HSD=10 MPa**

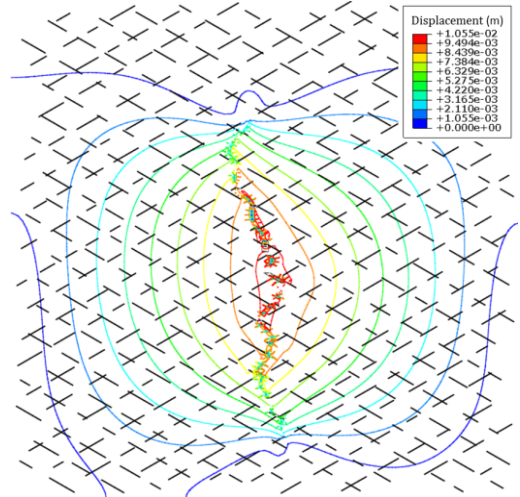

**Fig.7. Total displacement contour within the simulated domain at the injection time of 50 s, HSD=10 MPa**

Next, the horizontal stress difference is decreased to zero and now more complex fracture geometry is expected. **Fig.8** shows the simulated hydraulic fracture propagation path and the associated fracture width profile at different injection time. We can notice that the fracture geometry resembles an "elbow" shape and the two hydraulic fracture wings follow very different trajectories: the upper fracture wing propagates along the direction of one of the natural fracture sets after crossing a few weak planes, while the lower fracture wing propagates along a zig-zag path. It seems that it is easier for the upper fracture wing to propagate over time, especially after the propagating direction becomes aligned with one of the natural fracture set and reduces its propagation resistance. As such, the upper fracture wing receives more injection fluid, which leads to longer fracture half-length and large fracture width. **Fig.9** shows the total displacement contour within the simulated domain after 50 seconds of injection. It is not surprising that the displacement is larger close to the wellbore and smaller in the region that further from the wellbore.

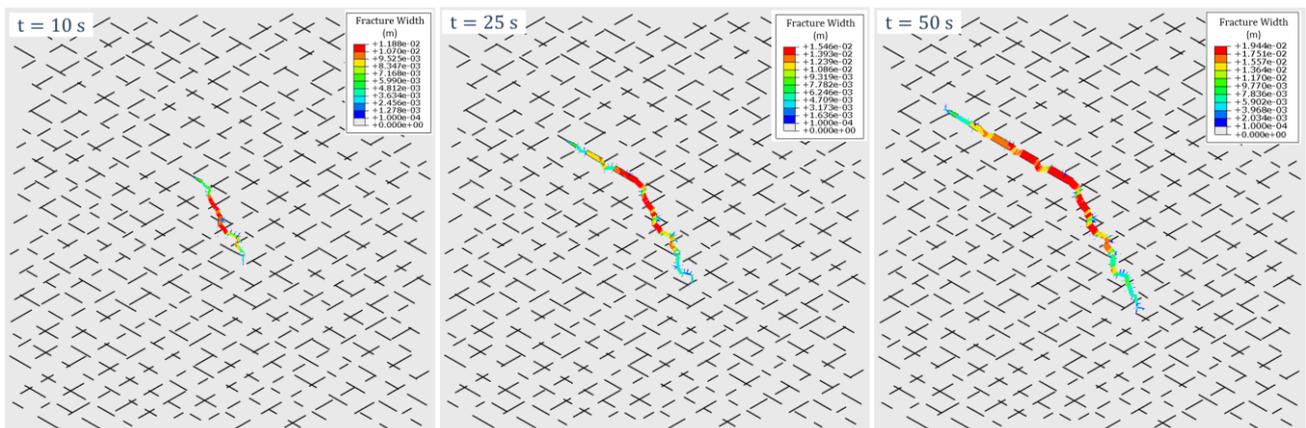

**Fig.8. Simulated hydraulic fracture propagation path and width profile at different injection time, HSD=0 MPa**

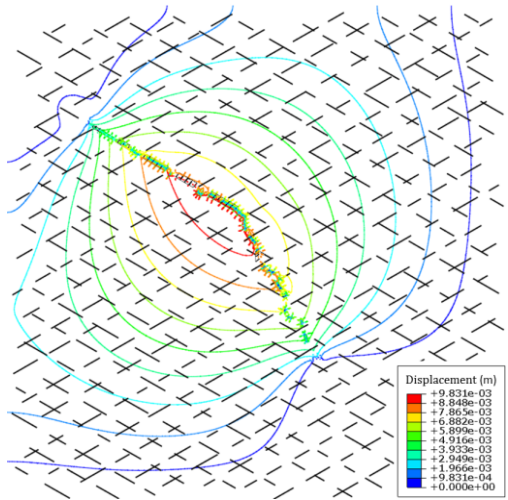

Fig.9. Total displacement contour within the simulated domain at the injection time of 50 s, HSD=0 MPa

In the previous simulations, it has demonstrated the scenarios where the horizontal stress difference is very large or absent, here we examine another case where the horizontal stress difference is moderate (i.e., HSD=5 MPa). **Fig.10** shows the simulated hydraulic fracture propagation path and the associated fracture width profile at different injection time. We can notice that the upper fracture wing was arrested after 10 seconds of injection, and because of this, most of the fluid flows into the lower fracture wing, and lower fracture wing hits the simulation boundary after 47 seconds of injection. In addition, the fracture width is smaller in the near-wellbore region, because of non-favorable fracture propagation direction that follows the natural fracture set direction, instead of open perpendicular to the minimum horizontal stress. Compared to the previous two cases, it is quite unexpected to witness that the zig-zag phenomenon is actually more pronounced when the horizontal stress difference is moderate. This stems from the fact that in this specific case, the horizontal stress difference is not large enough to dominate the propagation direction, but it is also not small enough to ensure fracture propagates along the direction of natural fracture sets. In other words, either force dictates the fracture propagation path under this scenario. This zig-zag fracture propagation manner poses substantial obstacles for proppant transport. A sharp diversion along the fracture propagation path can significantly reduce the proppant momentum when it hits the walls, which increases the risk of proppant blockage and premature screen-out. On top of that, this adverse impact can be exacerbated if the next section of fracture has a smaller fracture width when it is not well-aligned with the direction of maximum horizontal stress. **Fig.11** shows the total displacement contour within the simulated domain after 47 seconds of injection. It clearly reflects the fact the lower part of the simulation domain has large displacement because of the nature of non-symmetric fracture propagation.

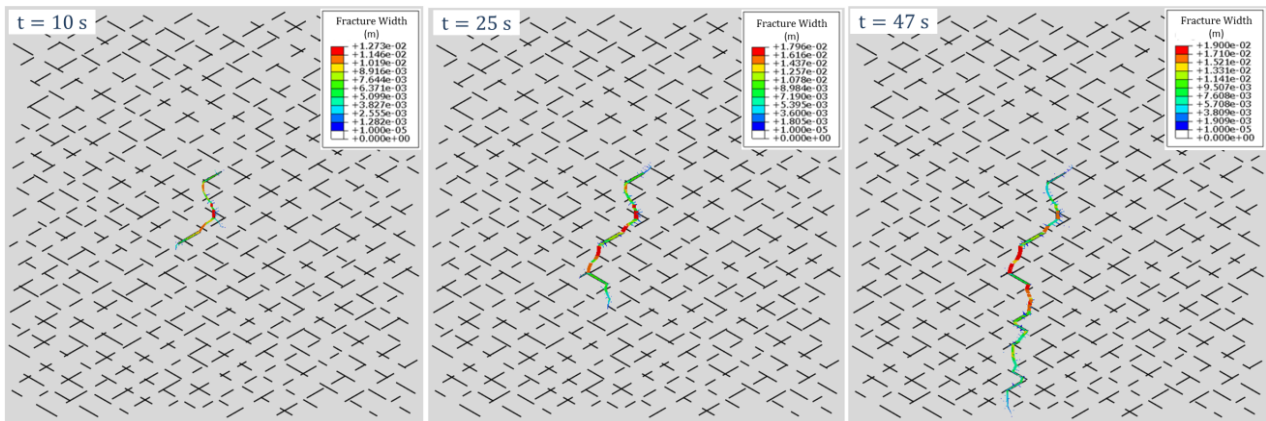

Fig.10. Simulated hydraulic fracture propagation path and width profile at different injection time, HSD=5 MPa

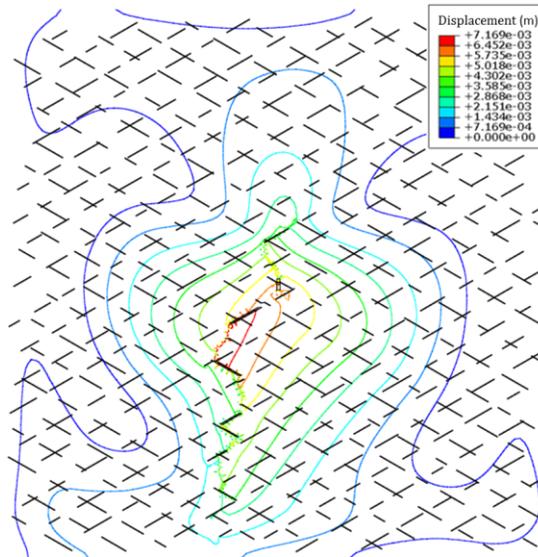

**Fig.11. Total displacement contour within the simulated domain at the injection time of 47 s, HSD=5 MPa**

Today, a widely accepted belief is that large stimulated reservoir volume (SRV) is crucial for stimulating unconventional reservoirs because of the complex fracture networks created within it, which forms inter-connected fracture flow path and extensively enhances reservoir contact area. And naturally fractured reservoirs with low horizontal stress difference are most likely to create complex fracture networks that sprawl all over the SRV during stimulation. However, the simulation results do not validate this common conception. Despite the fact that the simulated reservoir is naturally fractured and the hydraulic fracture geometry can be quite complex, but the so-called "fracture networks" are not observed for the presented cases. This begs the question: what mechanisms prevent the formation of complex fracture networks in naturally fractured reservoirs? And why doesn't hydraulic fracture propagate along all the branches and weak planes and form a treelike pattern?

The answer lies in stress shadow effect and resistance-dependent fluid distribution. Once the dominant propagating hydraulic fractures start to emerge, these dominant fracture channels inhibits the growth of nearby hydraulic fractures by pushing them towards an unfavorable direction, making it harder to propagate or forcing them to close because of increased local stress. In addition, fluid distribution among multiple propagating fractures is analogic to current flow in an electrical circuit network, the fluid amount a fracture receives is reversely proportional to its resistance (Yi and Sharma 2018), so the dominant hydraulic fractures receive the most injection fluid because of larger fracture width (with much less friction pressure drop along fracture due to higher conductivity) and less propagating resistance because of favorable propagation direction, which further hinders the growth of nearby fractures. In other words, once the advantage of the dominant hydraulic fractures is established, this advantage will be self-enhanced and the stress shadow effect will become even more severe as the dominant fracture channels become wider and longer. The above-presented case is a good manifestation of this phenomenon. If we take a closer look at the fracture propagation path at certain injection time frames, as shown in **Fig.12 (**The blue color represents open fracture with completely damaged cohesive element**)**, we can observe that the lower fracture wing has branched into two fractures after 4.3 seconds of injection, however, only the left branch continues to grow as it propagates along the direction of natural fracture set after 11.6 seconds of injection and the right branch gradually closes as the left branch become the dominant propagating fracture. The same situation occurs after 20.8 seconds of injection, when the lower wing hydraulic fracture branches along two natural fracture sets, however, the left branch ceases to grow as the right one becomes the main propagation channel and pushes the left branch to close.

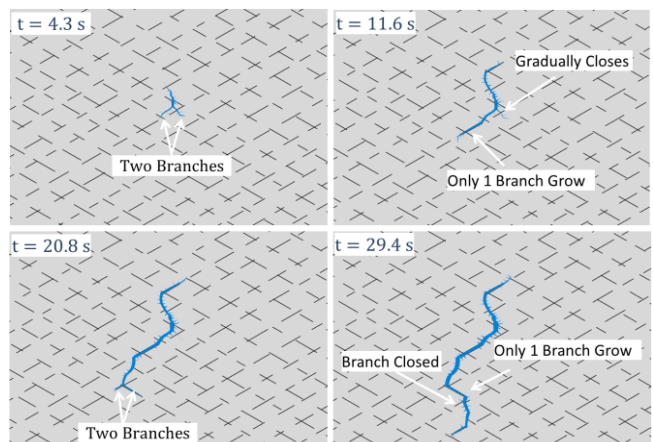

**Fig.12. Closer examination of simulated hydraulic fracture propagation path at different injection time, HSD=5 MPa**

One may argue that the absence of fracture networks in presented cases can result from the fact that the natural fracture bond is not weak enough. To verify this concept, the strength of the natural fracture bond is decreased 10 times, and to ensure a favorable condition for the development of fracture networks, the horizontal stress difference is set to be zero. **Fig.12** shows the simulated hydraulic fracture propagation path and the associated fracture width profile after 50 seconds of injection. Compared to Fig.8, we can conclude that the hydraulic fracture pattern is indeed, more complex when the natural fractures have a weaker cement bond. Now, the upper fracture wing has split into two main hydraulic fracture channels and because of stress show effect, they propagate away from each other. Nevertheless, this complex fracture and surrounding natural fracture do not form fracture networks.

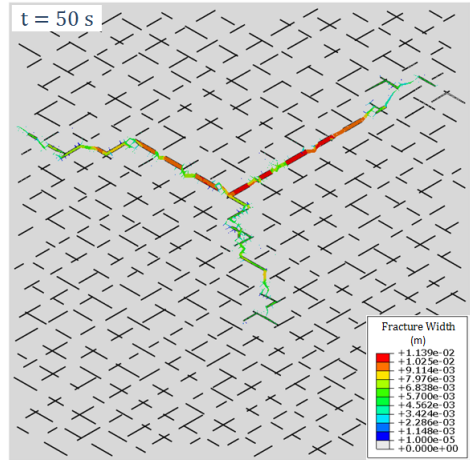

**Fig.13. Simulated hydraulic fracture propagation path and width profile at different injection time with weaker natural fracture bond, HSD=0 MPa**

A closer examination of the fracture propagation path, we can have a better idea of the dynamic fracture branch and closure behavior. For example, after 5.6 seconds of injection, the hydraulic fracture geometry is quite complex, as some fracture segments align with the natural fracture set while others propagate across it, as shown in **Fig.14**. Notice that shear failure can occur in the natural fracture at a distance away from the fracture tip. These shear failures of natural fractures discharge acoustic energy across the reservoir and can be detected via microseismic monitoring during stimulation. If these stimulated natural fractures are not well connected to the main hydraulic fracture, they have little contrition to production enhancement (Wang 2017). This also reflected in field observation that microseismic events do not correlate to production at all (Moos et al. 2011). After 13.4 seconds of injection, the complex fracture has evolved into a pattern that consists of three main fracture channels. Once the advantage of the dominant fractures is established, the trend of "Matthew Effect" is inevitable. Overall, the hydraulic fracture path looks more complex in the near-wellbore region than in the far field.

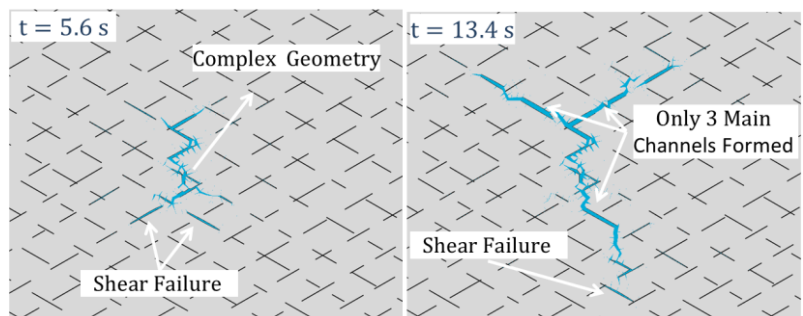

**Fig.14. Closer examination of simulated hydraulic fracture propagation path at different injection time with weaker natural fracture bond, HSD=0 MPa**

Next, maintain zero horizontal stress difference and increase rock Young's modulus from 10 GPa to 30 GPa, **Fig.15** shows the simulated hydraulic fracture propagation path and stress contour at different injection time. To better visualize hydraulic fracture branches and stress distribution, the grids that represent pre-existing natural fractures are removed. The stress along the x-axis is used as an indicator of the in-situ stress state (negative value designates compression and positive value designates tension). At 6.7 s injection time, it can be observed that the formation is more compressed near hydraulic fractures, and is under tensile state near the tip of hydraulic fracture branches and reactivated natural fractures. It is not surprising that the tip region of hydraulic fracture branches is always under tensile, as long as the fracture is open at the tip region. After 9.3 s of injection, hydraulic fractures continue to propagate and one can observe that much larger of simulation domain is under higher compression stress, except localized tensile stress at the tip of fracture branches. The propagation of hydraulic fractures create deformations in surrounding rocks and increases the compression stress due to the growth of the fracture width, this is how the stress shadow effect come into being, and each propagating hydraulic fracture branch will inevitably impact the growth of other adjacent hydraulic fracture branches due to this stress interference. From the simulation results, it seems that the hydraulic fracture is more complex and more hydraulic fracture branches develop during the propagation when Young's modulus is higher, especially close to the near-wellbore region. This can be attributed to the fact that both shear stress and stress shadow effect is elevated in the vicinity of propagating hydraulic fracture. Nevertheless, similar to previous case where Young's modulus is lower, three dominant hydraulic fracture paths eventually emerge and the secondary fracture branches cease to propagate after growing a short distance.

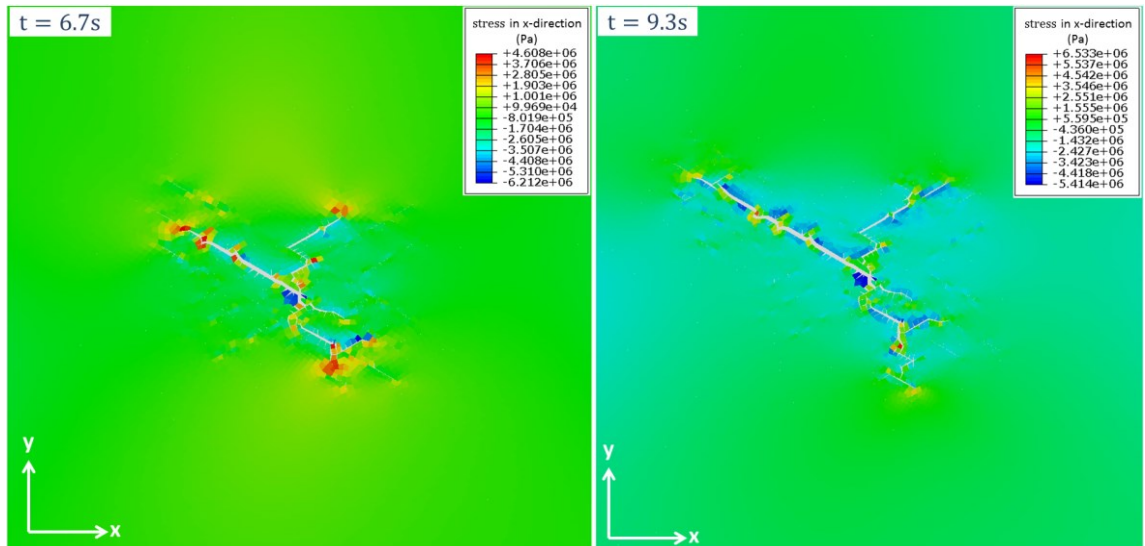

Fig.15. Simulated hydraulic fracture propagation path and stress at different injection time with weaker natural fracture bond, HSD=0 MPa, Young's modulus increases to 30 GPa

So, it can be concluded that in naturally fractured reservoirs, hydraulic fracture can turn, kink, branch and coalesce, and the fracture propagation path is complex, and is often asymmetric, but it does not necessary mean that the tree-like complex fracture networks, where different levels of fracture branches are well connected, can be formed. The main hydraulic fracture branches serve as the primary fluid-receiving channel and tend to propagate away from each other and the secondary fracture braches either cease to grow after propagation a short distance or gradually close because of stress shadow effect, fluid resistance and unfavorable propagation directions. Similar works using other numerical models (Chen et al. 2018; Li et al. 2017; Li et al. 2019; Ouchi et al. 2017; Xie et al. 2018) also suggest that only one or two dominant hydraulic fracture propagation paths exist per perforation cluster or fluid injection point in naturally fractured reservoirs and fracture branches are not creating intercepted networks. These are not just merely coincidence, other works (Profit et al. 2016; Wang et al. 2018; Zhao et al. 2014) that simulate hydraulic fracture propagation and induced seismicity in natural fractured reservoirs also observed that only one dominant hydraulic fracture propagation path exist, in spite of the fact that the simulated microseismic events cover a much larger reservoir volume. So, stimulation in naturally fractured reservoirs does not guarantee the creation of fracture networks inside SRV. Some studies (Cheng et al. 2017; Settgast et al. 2017) indeed demonstrate that open complex fracture networks can actually be formed around the main hydraulic fracture if the natural fractures are densely populated and well-connected with very weak cement bond, because hydraulic fracture does not need to break the intact rock to connect all the adjacent weak planes. Yet, laboratory experiments and fully coupled simulations (Kim et al 2017) indicate that it still can be the case where only one dominant hydraulic fracture channel emerges despite well-connected pre-existing fracture networks, even with low viscosity injection fluid, weak strength of pre-existing fractures. Daneshy (2019) presented an example of mine-back experiments, as shown in **Fig.16**. Similar to our presented case of Fig.6, it shows a roughly planar hydraulic fracture containing a green plastic proppant, crossing some natural fractures and deflecting along others. Presence of proppant indicates a continuous fracture, even though it appears segmented and discontinuous in this cross section. All these studies suggest that 'complex fracture', rather than 'fracture networks', are the norm in naturally fractured reservoirs, unless the pre-existing natural fractures or weak planes already formed 'networks' in the first place.

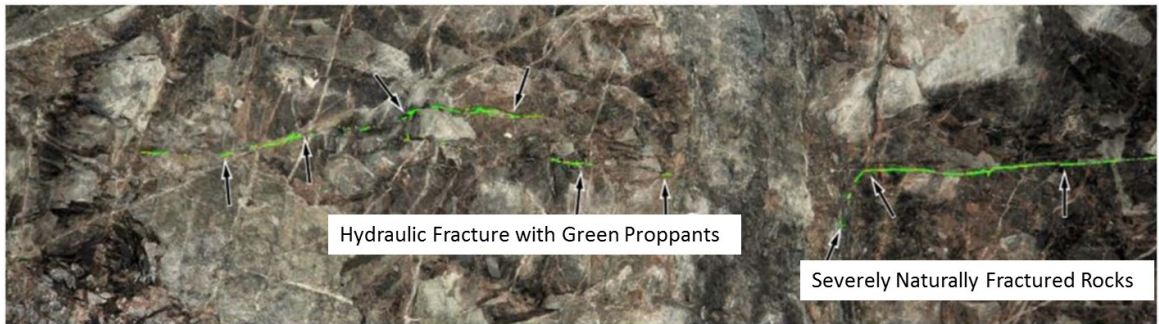
Fig.16. Hydraulic fracture in a severely naturally fractured formation (modified from Daneshy 2019)

## 5. Discussion

Modeling hydraulic fracture propagation in naturally fractured reservoirs is a challenging endeavor. The coupled multi-physics problem and the constant changes of boundaries make it difficult to solve the underlying partial differential equations with various imposing constraints. On top of that, some crucial data required for quantitative modeling and analysis are often difficult to obtain or have a lot of uncertainties. The most reliable method to estimate the minimum principal stress is through diagnostic fracture injection test or DFIT (Wang and Sharma 2017; 2019a). The maximum horizontal stress can be estimated via stress-induced wellbore breakouts if the minimum horizontal stress is known (Zoback 2007). The biggest challenge lies in delineating pre-existing natural fracture distribution pattern, density, azimuth, connectivity and the strength of the cement bond. While the mechanics and geologic conditions to generate natural fractures are generally well understood, the lack of hard data makes them difficult to describe in the subsurface. 3D seismic survey and ant tracking technology can be used for fault detection, but has limited application when it comes to smaller scale natural fracture set. Image logs and other downhole measurement can direct expose natural fractures, but it only reveals natural fractures that intercept the wellbore. And no practical method has been proposed to measure the properties of natural fracture cement bond under in-situ conditions. Even though pumping pressure, micro-seismic data and production data can be used to calibrate the fracture propagation model, but the substantial non-uniqueness of interpretation and associated uncertainties make it extremely hard for quantitative prediction. Furthermore, the lateral heterogeneity of natural fractures and reservoir properties pose additional challenges for mapping reservoir properties and fracture propagation modeling. Thus, how to calibrate hydraulic fracturing model by honoring the sparse input data with geologically based knowledge remains to be solved.

The concept of stimulated reservoir volume and associated fracture networks has been pivotal for hydraulic fracturing design in unconventional reservoirs. However, as many field observations indicate, the stimulated reservoir volume (identified by microseismic events) does not correlate to production and it is the drained reservoir volume (DRV) that really matters. How quickly the reservoir can be drained and what is the ultimate recovery hinges on the dimension, connectivity and conductivity of hydraulic fracture, but unfortunately, no consensus has been reached as regards to what the hydraulic fracture looks like in these naturally fractures reservoirs. **Fig. 17** illustrates three possible fracture typologies (complex fracture can have more than one main fracture channels if they diverge early on under low HSD condition). When the reservoir is homogenous and natural fractures are absent, hydraulic fracture will form a bi-wing planar fracture and propagates perpendicular to the direction of minimum principal stress. However, when abundant natural fractures exist, it completely changes the propagation behavior. Under this context, the hydraulic fracture can either form complex fracture geometry with a few dominant flow channels or create a complex fracture network where all the adjacent branches or weak planes are inter-connected. This study suggest that the former one is the most likely scenario, with some additional shear failed natural fractures that are not connected to the main hydraulic fracture if the natural fracture bond is weak.

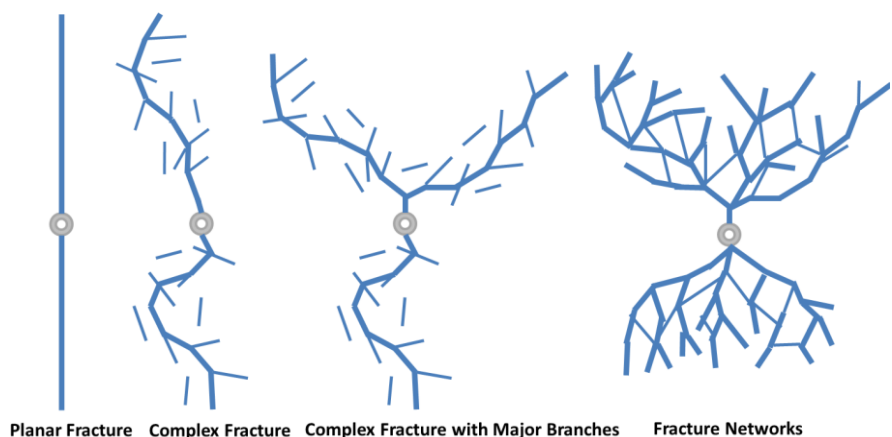
Fig.17. Illustration of possible hydraulic fracture typology

Even though shear failure is more likely to happen ahead of propagating hydraulic fracture tip when stress anisotropy is large (Wang et al. 2016), which may simulate more natural fractures, however, large stress anisotropy itself is favorable to fracture growth that perpendicular to the minimum principal stress, which, in return, constrains fracture complexity. Higher Young's modulus can also lead to the increase of shear stress in the vicinity of propagating hydraulic fracture and resulting in more stimulated natural fractures, but higher Young's modulus also intensifies the stress shadow effect that prohibits further fracture complexity. Once the tip of dominant hydraulic fracture channel passes the stimulated natural fractures or secondary fracture branches, the compression stress that resulted from the stress shadow effect will force these stimulated natural fractures or secondary fracture branches to close and cease to grow.

As a matter of fact, it is beneficial to make an analogy between multi-fracturing within a horizontal stage and the propagation of hydraulic branches, as shown in **Fig.18**. It is already well-known that stress shadow effect plays an important role in horizontal well fracturing. If the stress shadow effect from the previous hydraulic fracturing stage is not considered, the fracture growth form the middle perforation cluster can be prohibited during the pad injection period when the perforation cluster space is not large enough, and the fracture at the ends has the propensity to propagate away from each other. If we assume the pressure drop along the wellbore and perforation cluster is negligible, then the "multi-fracturing" scenario can be transformed into a case of fracture branching. Because the sheer scale of fracture branching is, at least, one order of magnitude smaller than simultaneously propagating fractures from different perforation clusters within a horizontal stage, the stress shadow effect is substantially larger between fracture branches than that between perforation clusters. If one or two of the fracture branches gain advantage at the very beginning caused by any random disturbance, it is most likely these branches grow to be the dominant ones, which not only less likely to be arrested by other natural fractures because of more energy it carries, but also prohibit the growth of other fracture branches.

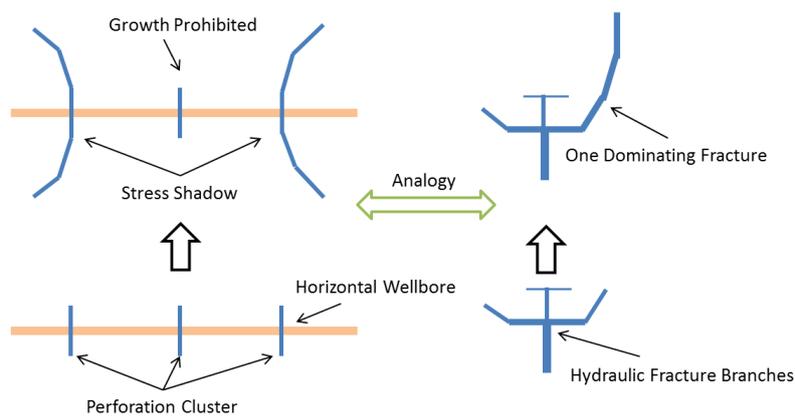

**Fig.18. The analogy between multi-fracturing within a certain stage and hydraulic fracture branches**

It has been argued that it is the "fracture networks" that created inside the SRV enhances the overall permeability, and this would justify larger cluster spacing to avoid stress shadow effect and the overlap of SRV. However, the trend in the industry towards tighter spacing (result in more productive wells) suggests that the creation of fracture networks inside SRV may not be the primary mechanism of production enhancement in many unconventional reservoirs. And in many cases, the production from optimized hydraulic fracture design using the concept of "fracture networks" underperforms the ones using only 3~4 m cluster spacing. Field observation also notice that planar fracture models with constant matrix permeability are capable of history matching production in most unconventional reservoirs, and rate transient analysis often reveals that productive surface area is much less than the "fracture networks" would imply. And the results from planar fracture propagation models match well with field data on inter-well pressure communication during hydraulic fracturing (Das et al., 2019; Seth et al., 2019). In addition, DFITs in most low permeability formations demonstrate extremely long after-closure linear flow behavior, which can only happen without the pressure interference of adjacent fractures. So, unless the "networks of pre-existing natural fractures" already exist in the first place, such as in coalbed methane (CBM) or some heavily fractured carbonate reservoirs, that exhibit highly stress-sensitive permeability in DFIT analysis (Wang and Sharma 2019a; 2019b) or rate transient analysis (Wang 2018), it seems that the "fracture network" may occur much less common than we expected in unconventional reservoirs.

Despite the fact that many numerical and small-scale laboratory studies have provided us valuable insight into the interactions between hydraulic fracture and natural fractures, comprehensive field studies of what complex hydraulic fracture really look like are urgently lacking. Raterman et al. (2017) presented a cross-study that involves coring multiple sidetracks directly through hydraulic fractures created by the stimulation of an adjacent horizontal well. This type of work is very rare and priceless, and may lead to a paradigm shift in how we envision hydraulic fracture inside SRV; such direct observations are the best benchmarks to validate our hypotheses about hydraulic fracturing in naturally fractured reservoirs. Some key aspects of their observations deserve emphasis: hydraulic fractures are discrete and have complex geometry, many of them are often accompanied by closely spaced, nearly paralleled fracture swarms. They saw no indication of flow in natural fractures in a number of orientations, and some hydraulic fractures can propagate over 450 m long. And microseismic events do not correlate to the density of natural fractures. If the formation of "fracture networks" is the norm in such fields, then it should be reflected in the coring samples. In addition, fracture networks can significantly reduce fracture length, but frac-hit caused by extremely long fractures that accidentally intercept nearby wells are commonplace in many naturally fractured shale plays. Thus, their observation seems in favor of "complex fracture", rather than "fracture networks" as depicted in Fig.17.

Currently, the occurrence of narrowly spaced (i.e on the scale of centimeters) fracture swarms that observed in the field is not captured in this presented model. Considering the strong stress shadow effect, these fractures swarms are most likely generated from unstable fracture propagation and branching, rather than propagate simultaneously all the way from the perforation clusters. Fracture propagation speed and rock heterogeneity are not the main reasons leading to unstable fracture propagation and branching, it is the dynamic flow of strain energy around the fracture tip and pile-up of elastic waves in the process zone controls the instability of fracture propagation (Bobaru and Zhang 2015). Eventually, one dominant fracture will emerge from these narrowly spaced fracture swarms (Wang 2016) and will propagate further by itself until the next moment of unstable fracture propagation occurs. This can possibly explain why fracture swarms are observed in some coring samples while not observed in others (Raterman et al. 2017). The advantage of using cohesive zone model is that the dynamic evolution of nucleation growth, and coalescence of microcracks occurs in the process zone ahead of fracture tip can be represented by cohesive zone, so that it is applicable to brittle, quasi-brittle and ductile rocks. However, just like using stress intensity factor to model fracture propagation, cohesive zone method is a local damage model and assumes quasi-static fracture propagation, so unstable fracture propagation and branching within intact rock is not accounted for. Non-local damage models are suitable candidates for modeling unstable fracture propagation, but it often requires significant computational efforts. So how much detail we need to model hydraulic fracture propagation is really a delicate balance among our modeling objectives, computational resources and the quality, resolution, uncertainty of the required input data available.

## 6. Conclusions

Many unconventional reservoirs are naturally fractured, and understanding the interactions between hydraulic fracture and natural fractures is crucial in estimating fracture complexity, drained reservoir volume and well stimulation efficiency. In this study, a fully coupled numerical model was developed to model hydraulic fracture process in naturally fractured reservoirs, where fracture turning, kinking, branching and coalescence can be captured. Simulation cases are presented including ones that contain pre-existing conjugate natural fracture sets. A comprehensive discussion of hydraulic fracture typology inside SRV and the implications of laboratory experiments and field observations are also presented. This study leads to the following conclusions:

1. There are two universal competing forces that govern the general trend of fracture propagation path in naturally fractured reservoirs: one is the propensity to propagate along the direction of the weak planes, the other is the tendency to propagate perpendicular to the direction of minimum principal stress. The orientations of natural fracture sets, the strength of natural fracture cement bond and in-situ stress anisotropy determine the overall direction of fracture propagation path.

2. Shear failure can happen along the natural fracture surface, even if these natural fractures are at a distance away from the propagating hydraulic fracture. But these shear-opened natural fractures are not necessarily connected to the hydraulic fracture.

3. High Young's modulus, weak natural fracture bond, and low-stress anisotropy promote fracture complexity. Fracture geometry tends to be more complex in the near-wellbore region than in the far-field.

4. Only a few dominant hydraulic fractures will emerge from multiple fracture branches and serve as the primary fluid-receiving channels. All the secondary fracture braches either cease to grow after propagation a short distance or gradually close because of stress shadow effect, flow-resistance dependent fluid distribution and unfavorable propagation directions. Thus, unless the existing natural fractures or weak planes already formed "fracture networks" in the first place, it is the asymmetrical, non-planar "complex fracture", rather than interconnected tree-like "fracture networks", developed inside stimulated reservoir volume of naturally fractured reservoirs during hydraulic fracturing.

Future work could include extending the presented model to 3D to examine the impact of spatial heterogeneous properties and plastic deformation on the interactions between hydraulic fracture and natural fractures